\documentclass[a4paper,11pt]{article}
\pdfoutput=1 % if your are submitting a pdflatex (i.e. if you have
\usepackage{jheppub} % for details on the use of the package, please
\usepackage[T1]{fontenc} % if needed

\usepackage{braket}

\title{\boldmath Direct Calculation of Mutual Information of Distant Regions} %Entanglement Entropy of Disjoint Regions in Excited States
\author{Noburo Shiba}

\affiliation{Theory Center, High Energy Accelerator Research Organization (KEK),\\ %Yukawa Institute for Theoretical Physics (YITP),\\
Tsukuba, Ibaraki 305-0801, Japan} %Kyoto University, Kyoto 606-8502, Japan}

\abstract{We consider the (R\'{e}nyi) mutual information, $I^{(n)}(A,B)=S^{(n)}_A+S^{(n)}_{B}-S^{(n)}_{A \cup B}$, of distant compact spatial regions A and B in the vacuum state of a free scalar field. 
The distance r between A and B is much greater than their sizes $R_{A,B}$. 
It is known that $I^{(n)}(A,B) \sim C^{(n)}_{AB} \bra{0} \phi(r)\phi(0) \ket{0}^2$ . 
We obtain the direct expression of $C^{(n)}_{AB}$ for arbitrary regions A and B. 
We perform the analytical continuation of $n$ and obtain the mutual information. 
The direct expression is useful for the numerical computation. 
By using the direct expression, we can compute directly $I(A,B)$ without computing $S_A, S_B$ and $S_{A \cup B}$ respectively, so it reduces significantly the amount of computation.

 }
 
\begin{document} 
	
\begin{flushright}
KEK-TH-2138
\end{flushright}

\maketitle
\flushbottom

\section{Introduction}
\label{intro}

The entanglement entropy in the quantum field theory plays 
important roles in many fields of physics including the string theory \cite{RT, Fa, Sw, NRT, MT, CMTU 2019,  Sh3, Sh4, Sh7, Sh8, Sh9, Sh10}, 
condensed matter physics \cite{LW, KP, CC, Sh11}, lattice gauge theories \cite{GST, Sh6}, cosmology \cite{Sh12}, and the physics of the black hole \cite{Bombelli:1986rw, Sr, SU, Ka, Sh1, Sh2}. 
The entanglement entropy is a useful quantity which characterizes quantum properties of given states. 
%For example, the entanglement entropy of ground states 
%follows the area law \cite{Bombelli:1986rw, Sr, Ereview, La} if we consider a local quantum field theory with a UV fixed point,
%while non-local field theories \cite{ShTa, Ka} or QFTs with fermi surfaces \cite{FS1} at UV cut off scale can violate the
%area law. 

For a given density matrix $\rho$ of the total system, 
the entanglement entropy of the subsystem $\Omega$ is defined as 
\begin{equation}
S_{\Omega} =-\mathrm{Tr} \rho_{\Omega} \ln \rho_{\Omega}, 
\end{equation}
where $\rho_{\Omega} =\mathrm{Tr}_{\Omega^{c}}\rho$ 
is the reduced density matrix of the subsystem $\Omega$ 
and $\Omega^c$ is the complement of $\Omega$. 
%The useful generalization of the entanglement entro
%$S_{\Omega}$ can be computed as limit as $n\rith$
The R\'{e}nyi entropy $S_{\Omega}^{(n)}$ is defined as 
\begin{equation}
S_{\Omega}^{(n)} = \dfrac{1}{1-n} \ln \mathrm{Tr} \rho_{\Omega}^{n} .
\end{equation}
The limit $n\rightarrow 1$ coincides with the entanglement entropy 
$\lim_{n=1} S_{\Omega}^{(n)}=S_{\Omega}$.

In this paper, we consider the (R\'{e}nyi) mutual information, $I^{(n)}(A,B)=S^{(n)}_A+S^{(n)}_{B}-S^{(n)}_{A \cup B}$, of distant compact spatial regions A and B in the vacuum state of a free scalar field. 
The distance r between A and B is much greater than their sizes $R_{A,B}$. 
It is known that \cite{Ca1}, when $r \gg R_{A,B} $, the (R\'{e}nyi) mutual information behaves as 
\begin{equation} \label{MI introduction}
	I^{(n)}(A,B) \sim C^{(n)}_{AB} \bra{0} \phi(r) \phi(0) \ket{0}^2 ,
\end{equation} 
%$I^{(n)}(A,B) \sim C^{(n)}_{AB} \left< \phi(r)\phi(0) \right>^2$ . 
where $C^{(n)}_{AB}$ depends on the shapes of the regions A and B.  
When both A and B are the spheres and the scalar field is massless, the coefficient $C^{(n)}_{AB}$ was calculated analytically by Cardy \cite{Ca1}.
However, it is difficult to calculate $C^{(n)}_{AB}$ analytically when both A and B are not the spheres or the scalar field is not massless. 
In this paper, we obtain the direct expression of $C^{(n)}_{AB}$ for arbitrary regions A and B in the vacuum state of a scalar field which has a general dispersion relation. 
We perform the analytical continuation of $n$ and obtain the mutual information $I(A,B)=\lim_{n \to 1} I^{(n)}(A,B)$. 
The direct expression is useful for the numerical computation. 
By using the direct expression, we can compute directly $I(A,B)$ without computing $S_A, S_B$ and $S_{A \cup B}$ respectively, so it reduces significantly the amount of computation.

We comment on the advantages of this direct expression over the conventional numerical computation by the real time formalism. 
Entanglement entropy in free scalar fields can be calculated numerically by the real time formalism \cite{Bombelli:1986rw,Sr}.
In order to calculate the coefficient $C^{(n=1)}_{AB}$ by the real time formalism, we have to plot the mutual information $I(A,B)$ as a function of r and extract the coefficient \cite{Sh2}.
So we have to calculate numerically $S_{A\cup B}$ many times to plot $I(A,B)$ as a function of r.
On the other hand, in our method, we separate the r dependence of $I(A,B)$ analytically and obtain the direct expression of $C^{(n=1)}_{AB}$.
So, it reduces significantly the amount of computation.

To obtain the direct expression of $C^{(n)}_{AB}$, 
we use the operator method to compute the R\'{e}nyi entropy developed in \cite{Sh5}. 
This operator method is based on the idea that 
$\mathrm{Tr} \rho_{\Omega}^n$ 
%the trace of the $n$th power of the reduced density matrix 
is written  
%of the subsystem 
as the expectation value of the local operator at $\Omega$. 
This idea was originally used to compute $I^{(n)}(A,B)$ in the vacuum state 
by Cardy \cite{Ca1}, Calabrese et al. \cite{Ca2}  and Headrick \cite{He}. 
%We generalize 
This idea was generalized to an arbitrary density matrix $\rho$ and the local operator was explicitly constructed in \cite{Sh5}. 
 Cardy's work \cite{Ca1} was generalized (at least for the computation of the mutual information as opposed to the mutual R\'{e}nyi information) for any CFT with a scalar in \cite{AF 2016}.
 The next to leading terms in the long distance expansion of the mutual information in a free scalar theory was studied in \cite{ACS 2016}. 
 The leading term for the mutual R\'{e}nyi information for two widely separated identical compound systems in a  free scalar theory was studied in \cite{Sc 2014}.

The present paper is organized as follows. 
In section 2, we review the operator method to compute the R\'{e}nyi entropy developed in \cite{Sh5}.
In section 3, we expand the glueing operator which plays the important role in the operator method to compute the (R\'{e}nyi) mutual information.
In section 4, we compute the (R\'{e}nyi) mutual information and obtain the direct expression of $C^{(n)}_{AB}$.

\section{The review of the operator method to compute the R\'{e}nyi entropy}
We review the operator method to compute the R\'{e}nyi entropy developed in \cite{Sh5}.
%$\mathrm{Tr} \rho_{\Omega}^n$ is expressed as follows. 
We consider $n$ copies of the scalar fields in (d+1) dimensional spacetime and the $j$-th copy of the scalar field is denoted by 
$\{ \phi^{(j)} \}$. 
Thus the total Hilbert space, $H^{(n)}$, is the tensor product of the $n$ copies of the Hilbert space, 
$H^{(n)}= H \otimes H \dots \otimes H$ where $H$ is the Hilbert space of one scalar field. 
We define the density matrix $\rho^{(n)}$ in  $H^{(n)}$ as 
\begin{equation}
\rho^{(n)} \equiv \rho \otimes \rho \otimes \dots \otimes \rho
\end{equation}
where $\rho$ is an arbitrary density matrix in $H$. 
We can express $\mathrm{Tr} \rho_{\Omega}^n$ as 
\begin{equation}
\mathrm{Tr} \rho_{\Omega}^n=\mathrm{Tr} (\rho^{(n)} E_{\Omega}),  \label{expectation intro}
\end{equation}
where 
\begin{equation} 
\begin{split}
&E_{\Omega} =\int \prod_{j=1}^{n} \prod_{x \in \Omega} DJ^{(j)}(x) D K^{(j)}(x)  
\exp [i \int d^d x \sum_{l=1}^{n} J^{(l+1)}(x) \phi^{(l)}(x) ] \\
&\times  \exp [i\int d^d x \sum_{l=1}^{n} K^{(l)}(x) \pi^{(l)}(x) ]
\times \exp [-i\int d^d x \sum_{l=1}^{n} J^{(l)} \phi^{(l)} ] ,
\end{split}  \label{opjkf intro}
\end{equation}
where $\pi(x)$ is a conjugate momenta of $\phi(x)$, $[\phi(x), \pi(y)]=i \delta^d (x-y)$, 
and $J^{(j)}(x)$ and $K^{(j)}(x)$ exist only in $\Omega$ and $J^{(n+1)}=J^{(1)}$ and 
we normalize the measure of the functional integral as 
$\int DJ^{(j) } \exp [i\int d^d x J^{(j)} (x) f(x) ] =\prod_{x \in \Omega} \delta (f(x)) $ 
where $f(x)$ is an arbitrary function. 
Notice that $\phi$ and $\pi$ in (\ref{opjkf intro}) are operators and the ordering is important. 
%We call 
This operator $E_{\Omega}$ is called as  \textit{the glueing operator}. 
When $\rho$ is a pure state, $\rho=\ket{\Psi}\bra{\Psi}$, 
the equation (\ref{expectation intro}) becomes 
\begin{equation}
\mathrm{Tr} \rho_{\Omega}^n = \bra{\Psi^{(n)}} E_{\Omega} \ket{\Psi^{(n)}}  \label{formula intro}
\end{equation}
where 
\begin{equation}
\ket{\Psi^{(n)}} =\ket{\Psi}\ket{\Psi}\dots \ket{\Psi} .
\end{equation}
%We will obtain the explicit form (\ref{opjkf intro}) of $E_{\Omega}$ and 
%investigate the general properties of  $E_{\Omega}$ in Section 2. 
%For a free scalar field, we can rewrite $E_{\Omega}$ in (\ref{opjkf intro}) 
%using the normal ordering. 

The useful property of the glueing operator for calculating the mutual information is the locality.
When $\Omega=A\cup B$ and $A \cap B=\emptyset$, 
\begin{equation}
E_{A\cup B} = E_{A} E_{B}   . \label{property2}
\end{equation}
From the locality (\ref{property2}), the mutual R\'{e}nyi information in the vacuum state can be expressed as the correlation function of the glueing operators,
\begin{equation} \label{locality 2}
\frac{\mathrm{Tr} \rho_{A\cup B}^{(n)}}{\mathrm{Tr} \rho_{A}^{(n)} \mathrm{Tr} \rho_{B}^{(n)}} = \frac{ \bra{0^{(n)}} E_{A} E_{B} \ket{0^{(n)}}  }{\bra{0^{(n)}} E_{A} \ket{0^{(n)}} \bra{0^{(n)}} E_{B}  \ket{0^{(n)}} } .
\end{equation}

We consider  $(d+1)$ dimensional free scalar field theory. 
For free scalar fields, it is useful to represent the glueing operator $E_{\Omega}$ 
in (\ref{opjkf intro}) as the normal ordered operator. 
We decompose $\phi$ and $\pi$ into the creation and annihilation parts, 
\begin{equation}
\phi (x) =\phi^{+}(x) +\phi^{-}(x), ~~~\pi (x) =\pi^{+}(x) +\pi^{-}(x),
\end{equation}
where 
\begin{equation}
\begin{split}
&\phi^{+} (x) = \int \dfrac{d^d p}{(2\pi)^d} \dfrac{1}{\sqrt{2E_p}} a_p e^{ipx}, 
~~\phi^{-} (x)=(\phi^{+} (x))^{\dagger} ,  \\
&\pi^{+} (x) = \int \dfrac{d^d p}{(2\pi)^d} (-i) \sqrt{\dfrac{E_p}{2}} a_p e^{ipx}, 
~~\pi^{-} (x)=(\pi^{+} (x))^{\dagger}  ,
\end{split}
\end{equation}
here  $E_p$ is the energy and  $[a_p,a_{p'}^{\dagger}] =(2\pi)^d \delta^{d} (p-p')$. 
The commutators of these operators are 
\begin{equation}
\begin{split}
&[\phi^{+} (x), \phi^{-} (y)]  = \bra{0} \phi (x) \phi (y) \ket{0} 
= \int \dfrac{d^d p}{(2\pi)^d} \dfrac{1}{2E_p}  e^{ip(x-y)} \equiv \dfrac{1}{2} W^{-1}(x,y),   \\
&[\pi^{+} (x), \pi^{-} (y)]  = \bra{0} \pi (x) \pi (y) \ket{0} 
= \int \dfrac{d^d p}{(2\pi)^d} \dfrac{E_p}{2}  e^{ip(x-y)} \equiv \dfrac{1}{2} W(x,y),          \\
&[\pi^{+} (x), \phi^{-} (y)] =[\pi^{-} (x), \phi^{+} (y)] = -\dfrac{i}{2} \delta^d (x-y)  ,
\end{split}  \label{commutators}
\end{equation}
where 
we have defined the matrices $W$ and $W^{-1}$ which has continuous indices $x,y$ in (\ref{commutators})
and $W^{-1}$ is the inverse of $W$. 
$W$ and $W^{-1}$ are positive definite symmetric matrices. 
By using (\ref{commutators}) and the Baker-Campbell-Hausdorff (BCH) formula 
$e^X e^Y =e^{[X,Y]} e^Y e^X, ~~ e^{X+Y} = e^{-\frac{1}{2}[X,Y] } e^X e^Y$,  
for $[[X, Y], X] = [[X, Y], Y]=0$, 
we obtain 
\begin{equation}
\begin{split}
&\exp [i\int d^d x J' \phi ] \exp [i\int d^d x K \pi ]  \exp [-i\int d^d x J \phi ]  \\
&=:\exp [i\int d^d x (K \pi + (J'-J)\phi)] :   \\
&\times \exp[\int d^d x d^d y (-\dfrac{1}{4} K(x) A(x,y)K(y) -\dfrac{1}{4} (J'-J)(x) D(x,y)(J'-J)(y) )  \\
&-\int d^dx \dfrac{i}{2} K(x) (J'+J)(x)  ] ,  \\
\end{split}  \label{normal order}
\end{equation}
where $:O:$ means the normal ordered operator of $O$. 
From (\ref{normal order}) 
we can rewrite $E_{\Omega}$  in (\ref{opjkf intro}) as 
the normal ordered operator,  
\begin{equation}
\begin{split}
&E_{\Omega} =\int \prod_{j=1}^{n} \prod_{x \in \Omega} DJ^{(j)}(x) D K^{(j)}(x)  
:\exp [i \sum_{l=1}^{n} \int d^d x ( (J^{(l+1)} -J^{(l)} ) \phi^{(l)} +K^{(l)} \pi^{(l)} ) ]: \exp [- \tilde{S}] , \\
\end{split}  \label{normal opjkf}
\end{equation}
where $J^{(n+1)} =J^{(1)}$ and %$K^{(n+1)} =K^{(1)}$ and  
\begin{equation}
\begin{split}
& \tilde{S} \equiv  \sum_{l=1}^{n} [\int d^d x d^d y [\dfrac{1}{4} K^{(l)}(x) W(x,y) K^{(l)}(y) 
+\dfrac{1}{4} (J^{(l+1)}- J^{(l)} ) (x) W^{-1}(x,y) (J^{(l+1)}- J^{(l)} )(y)]  \\
&+\dfrac{i}{2} \int d^d x K^{(l)}(x) (J^{(l+1)}+J^{(l)})(x) ] . 
\end{split}  \label{tilde S}
\end{equation}

\section{The expansion of the glueing operator}
We consider a complex scalar field $\phi$ because it is useful for later calculation. 
The mutual information of a real free scalar field can be obtained by dividing the  mutual information of the complex free scalar field by 2.
Then, the glueing operator becomes
\begin{equation}
\begin{split}
E_{\Omega} =&\int \prod_{j=0}^{n-1} \prod_{x \in \Omega} DJ^{(j)}(x) D K^{(j)}(x)  
:\exp [i \sum_{l=0}^{n-1} \int d^d x ( (J^{(l+1)} -J^{(l)} ) \phi^{(l)*} +K^{(l)} \pi^{(l)*}\\
&+ (J^{(l+1)*} -J^{(l)*} ) \phi^{(l)} +K^{(l)*} \pi^{(l)} ) ]: \exp [- \tilde{S}] , \\
\end{split}  \label{normal opjkf complex}
\end{equation}
where
\begin{equation}
\begin{split}
 \tilde{S} \equiv&  \sum_{l=1}^{n} [\int d^d x d^d y [\dfrac{1}{2} K^{(l)}(x) A(x,y) K^{(l)*}(y) 
+\dfrac{1}{2} (J^{(l+1)}- J^{(l)} ) (x) D(x,y) (J^{(l+1)*}- J^{(l)*} )(y)]  \\
&+\dfrac{i}{2} \int d^d x(K^{(l)}(x) (J^{(l+1)*}+J^{(l)*})(x)+K^{(l)*}(x) (J^{(l+1)}+J^{(l)})(x)) ] . 
\end{split}  \label{tilde S complex}
\end{equation}
For the free scalar field, it is useful to use the following Fourier transformation,
\begin{equation} \label{Fourier}
\begin{split}
f^{(l)}=\frac{1}{\sqrt{n}} \sum_{k=0}^{n-1} e^{2 \pi i k l/n} \tilde{f}^{(k)}
\end{split} 
\end{equation}
where $f^{(l)}$ is an arbitrary n dimensional vector and $ \tilde{f}^{(k)}$ is its Fourier transformation, i.e. (\ref{Fourier}) is the definition of the Fourier transformation.
The Fourier transformation diagonalizes the glueing operator, 
\begin{equation} \label{Fourier E}
\begin{split}
E_{\Omega} =\prod_{k=0}^{n-1} E_{\Omega}^{(k)}
\end{split}  
\end{equation}
where
\begin{equation}
\begin{split}
E_{\Omega}^{(k)} =\int \prod_{x\in \Omega}  D\tilde{J}^{(k)}(x) D \tilde{K}^{(k)}(x) :\exp[i Q^{(k)}] : \exp[-\tilde{S}^{(k)}]
\end{split}  
\end{equation}
\begin{equation}
\begin{split}
 Q^{(k)}\equiv \int d^dx [ (e^{2 \pi ik/n} -1) \tilde{J}^{(k)} \tilde{\phi}^{(k)*} +(e^{-2 \pi ik/n} -1) \tilde{J}^{(k)*} \tilde{\phi}^{(k)} +\tilde{K}^{(k)} \tilde{\pi}^{(k)*}+\tilde{K}^{(k)*} \tilde{\pi}^{(k)} ]
\end{split}  
\end{equation}
\begin{equation}
\begin{split}
 \tilde{S}^{(k)} \equiv& \int d^d x d^d y [\dfrac{1}{2} \tilde{K}^{(k)}(x) A(x,y) \tilde{K}^{(k)*}(y) 
+\dfrac{1}{2} (1-\cos (\frac{2\pi k}{n})) \tilde{J}^{(k)}  (x) D(x,y)  \tilde{J}^{(k)*} (y)]  \\
&+\dfrac{i}{2} \int d^d x( (e^{-2\pi i k/n} +1  ) \tilde{K}^{(k)}(x) \tilde{J}^{(k)*}(x)+ (e^{2\pi i k/n} +1  ) \tilde{K}^{(k)*}(x) \tilde{J}^{(k)}(x))  . 
\end{split}  \label{tilde S complex k}
\end{equation}
In order to expand $:\exp[i Q^{(k)}] :$ in $E_{\Omega}^{(k)}$, 
we define $\langle \dots \rangle $ as 
\begin{equation}
\begin{split}
&\langle \dots \rangle \equiv \frac{ \int \prod_{x\in \Omega}  D\tilde{J}^{(k)}(x) D \tilde{K}^{(k)}(x)  \exp[-\tilde{S}^{(k)}] \dots }{\int \prod_{x\in \Omega}  D\tilde{J}^{(k)}(x) D \tilde{K}^{(k)}(x)  \exp[-\tilde{S}^{(k)}]  } 
\end{split}  
\end{equation}
where $\dots$ is an arbitrary function of $\tilde{J}^{(k)}$ and $ \tilde{K}^{(k)}$.
When $\Omega$ is a compact spatial region, we express $E_{\Omega}^{(k)}$ as a sum of the local operators at a conventionally chosen point $x_0$ inside $\Omega$.
Thus, we expand $E_{\Omega}^{(k)}$ as 
\begin{equation}
\begin{split}
&\frac{E_{\Omega}^{(k)}}{\bra{0} E_{\Omega}^{(k)} \ket{0} } = 1- \frac{1}{2} \langle : ( Q^{(k)} )^2  : \rangle + \cdots  \\
&= 1- \left(2-2 \cos \left( \frac{2\pi k}{n} \right) \right) \int d^d x d^d y \langle \tilde{J}^{(k) }  (x) \tilde{J}^{(k) *}  (y) \rangle : \tilde{\phi}^{(k)*}(x) \tilde{\phi}^{(k)}(y): + \cdots \\
&= 1-: \tilde{\phi}^{(k)*}(x_0) \tilde{\phi}^{(k)}(x_0): \left(2-2 \cos \left( \frac{2\pi k}{n} \right) \right) \int d^d x d^d y \langle \tilde{J}^{(k) }  (x) \tilde{J}^{(k) *}  (y) \rangle  + \cdots .
\end{split}  
\end{equation}
In order to represent the Gauss integrals of  $\tilde{K}^{(k)}$ and  $\tilde{J}^{(k)}$, 
we will use the following matrix notation, 
\begin{alignat}{2} 
W(x,y) = \begin{pmatrix} 
W(x_\Omega , y_\Omega) & W(x_\Omega , y_{\Omega^c})  \\
W(x_{\Omega^c} , y_\Omega) & W(x_{\Omega^c} , y_{\Omega^c}) 
\end{pmatrix} 
\equiv   \begin{pmatrix} 
A & B  \\
B^T & C 
\end{pmatrix}  &   ~~~~~~ \label{definition A} \\ 
W^{-1}(x,y) = \begin{pmatrix} 
W^{-1}(x_\Omega , y_\Omega) & W^{-1}(x_\Omega , y_{\Omega^c})  \\
W^{-1}(x_{\Omega^c} , y_\Omega) & W^{-1}(x_{\Omega^c} , y_{\Omega^c}) \end{pmatrix}
\equiv   \begin{pmatrix} 
D & E  \\
E^T & F 
\end{pmatrix}      \label{definition D}       
%\label{eq:1-4}
\end{alignat}
where $x_{\Omega (\Omega^c)}$ and $y_{\Omega (\Omega^c)}$ are the coordinates 
in $\Omega (\Omega^c)$, where $\Omega^c$ is the complement of $\Omega$.

In order to calculate $ \langle \tilde{J}^{(k) }  (x) \tilde{J}^{(k) *}  (y) \rangle$, we perform the $\tilde{K}^{(k)}$ integral first,
\begin{equation} \label{K integral}
\begin{split}
&\int \prod_{x\in \Omega}  D \tilde{K}^{(k)}(x) \exp[-\tilde{S}^{(k)}]  \\
&= \frac{1}{ \det \left( \frac{A}{2 \pi} \right)} 
\exp [ - \int d^d x d^d y \tilde{J}^{(k) *}  (x) \left( A^{-1} +D+ \cos \left( \frac{2\pi k}{n} \right) (A^{-1} -D ) \right) (x,y) \tilde{J}^{(k)}  (x) ]. 
\end{split}  
\end{equation}
From (\ref{K integral}), we obtain
\begin{equation}
\begin{split}
&\langle \tilde{J}^{(k) }  (x) \tilde{J}^{(k) *}  (y) \rangle 
= \frac{ \int \prod_{x\in \Omega}  D\tilde{J}^{(k)}(x) D \tilde{K}^{(k)}(x)  \exp[-\tilde{S}^{(k)}] \tilde{J}^{(k) }  (x) \tilde{J}^{(k) *}  (y) }{\int \prod_{x\in \Omega}  D\tilde{J}^{(k)}(x) D \tilde{K}^{(k)}(x)  \exp[-\tilde{S}^{(k)}]  } \\
&= \left( A^{-1} +D+ \cos \left( \frac{2\pi k}{n} \right) (A^{-1} -D ) \right)^{-1} (x, y).
\end{split}  
\end{equation}
In order to separate the n dependence of $ \langle \tilde{J}^{(k) }  (x) \tilde{J}^{(k) *}  (y) \rangle$, 
we rewrite it as
\begin{equation}
\begin{split}
& \left( A^{-1} +D+ \cos \left( \frac{2\pi k}{n} \right) (A^{-1} -D ) \right)^{-1}=X \left(1- \cos \left( \frac{2\pi k}{n} \right)Y \right)^{-1} X
\end{split}  
\end{equation}
where
\begin{equation} \label{definition X Y}
\begin{split}
& X \equiv (A^{-1}+D )^{-1/2}, \; \; \;   Y\equiv X (D-A^{-1}) X
\end{split}  
\end{equation}
\begin{equation} \label{definition lambda}
\begin{split}
& Y=O^{T} \Lambda O , \;\;\; \Lambda =\mathrm{ diag} (\lambda_i)
\end{split}  
\end{equation}
Thus we obtain
\begin{equation}
\begin{split}
& \left( X \left(1- \cos \left( \frac{2\pi k}{n} \right)Y \right)^{-1} X  \right)_{i,j}= \sum_{l} Z_{li} 
\frac{1}{1-\lambda_l \cos  \left( \frac{2\pi k}{n} \right)  } Z_{lj}
\end{split}  
\end{equation}
where $Z=OX$ and we discretized the space coordinates in order to regularize the scalar field.   
In the appendix \ref{lambda range proof}, we show that the range of the eigenvalues $\lambda_i$ is
\begin{equation} \label{lambda range}
\begin{split}
0 \leq \lambda_i < 1 .
\end{split}  
\end{equation}
Finally, when $\Omega$ is a compact spatial region, 
we obtain the expansion of $E_{\Omega}^{(k)}$ as
\begin{equation} \label{Expansion result}
\begin{split}
&\frac{E_{\Omega}^{(k)}}{\bra{0} E_{\Omega}^{(k)} \ket{0} } = 1-: \tilde{\phi}^{(k)*}(x_0) \tilde{\phi}^{(k)}(x_0): C_{\Omega}^{(k)} + \cdots ,
\end{split}  
\end{equation}
where
\begin{equation} \label{C_Omega}
\begin{split}
& C_{\Omega}^{(k)}\equiv \left(2-2 \cos \left( \frac{2\pi k}{n} \right) \right) \int d^d x d^d y \langle \tilde{J}^{(k) }  (x) \tilde{J}^{(k) *}  (y) \rangle  \\
&= \left(2-2 \cos \left( \frac{2\pi k}{n} \right) \right) \sum_{i} \sum_{j} \sum_{l} Z_{l i} \frac{1}{1-\lambda_{l} \cos  \left( \frac{2\pi k}{n} \right)  } Z_{l j} \\
&\equiv \left(2-2 \cos \left( \frac{2\pi k}{n} \right) \right) \sum_{i_\Omega} \sum_{j_\Omega} \sum_{l_\Omega} Z^{(\Omega)}_{l_\Omega i_\Omega} \frac{1}{1-\lambda^{(\Omega)}_{l_{\Omega}} \cos  \left( \frac{2\pi k}{n} \right)  } Z^{(\Omega)}_{l_\Omega j_\Omega} .
\end{split}  
\end{equation}
In the last line in (\ref{C_Omega}), we added the subscript and the superscript $\Omega$ in order to clarify that $i_\Omega$, $j_\Omega$ and $l_\Omega$ are the coordinates in $\Omega$, and  $Z^{(\Omega)}$ and $\lambda^{(\Omega)}$ depend on $\Omega$.

\section{The (R\'{e}nyi) mutual information of distant regions}
We apply above results to the mutual R\'{e}nyi information $I^{(n)}(A,B)$ of disjoint
compact spatial regions A and B in the vacuum states of the free scalar field.
From (\ref{locality 2}), (\ref{Fourier E}) and (\ref{Expansion result}), we obtain
\begin{equation} \label{Expansion trace}
\begin{split}
& \frac{\mathrm{Tr} \rho_{A\cup B}^{n} }{\mathrm{Tr} \rho_{A}^{n} \mathrm{Tr} \rho_{ B}^{n} } 
= \prod_{k=0}^{n-1} \frac{ \bra{0} E_{A}^{(k)} E_{B}^{(k)} \ket{0}  }{\bra{0} E_{A}^{(k)} \ket{0} \bra{0} E_{B}^{(k)}  \ket{0} } \\
&\simeq \prod_{k=0}^{n-1} \bra{0} (1-: \tilde{\phi}^{(k)*}(x_A) \tilde{\phi}^{(k)}(x_A): C_{A}^{(k)} ) (1-: \tilde{\phi}^{(k)*}(x_B) \tilde{\phi}^{(k)}(x_B): C_{B}^{(k)} ) \ket{0} \\
&= \prod_{k=0}^{n-1}  (1+ C_{A}^{(k)} C_{B}^{(k)} f(r)  )
\end{split}  
\end{equation}
where $x_A$ and $x_B$ are some conventionally chosen points inside A and B,  $r=|x_A -x_B|$, and 
\begin{equation}
\begin{split}
f(r) &\equiv \bra{0} : \tilde{\phi}^{(k)*}(x_A) \tilde{\phi}^{(k)}(x_A): : \tilde{\phi}^{(k)*}(x_B) \tilde{\phi}^{(k)}(x_B):  ) \ket{0} \\
&= \left( \bra{0} \phi(x_A) \phi^{*}(x_B) \ket{0} \right)^2
\end{split}  
\end{equation}
From (\ref{Expansion trace}), we obtain the mutual R\'{e}nyi information as
\begin{equation} \label{Renyi 1}
\begin{split}
&I^{(n)} (A, B) = \frac{1}{n-1} \ln \frac{\mathrm{Tr} \rho_{A\cup B}^{n} }{\mathrm{Tr} \rho_{A}^{n} \mathrm{Tr} \rho_{ B}^{n} } \simeq \frac{1}{n-1} \sum_{k=0}^{n-1} \ln \left( 1+  C_{A}^{(k)}  C_{B}^{(k)} f(r)  \right)
 \simeq \frac{ f(r)}{n-1} \sum_{k=0}^{n-1}  C_{A}^{(k)}  C_{B}^{(k)}
\end{split}  
\end{equation}
We substitute $ C_{\Omega}^{(k)}$ in (\ref{C_Omega}) into (\ref{Renyi 1}) and obtain 
\begin{equation} \label{MI final}
\begin{split}
&I^{(n)} (A, B) 
\simeq  C_{AB}^{(n)}   f(r) ,      
\end{split}  
\end{equation}
\begin{equation} \label{C_{AB}}
\begin{split}
&C^{(n)}_{AB}
\equiv \frac{4 }{n-1}   \sum_{i_A} \sum_{j_A} \sum_{l_A} \sum_{i_B} \sum_{j_B} \sum_{l_B} 
 Z_{l_A i_A}^{(A)} Z_{l_A j_A}^{(A)} Z_{l_B i_B}^{(B)} Z_{l_B j_B}^{(B)} F(n, \lambda_{l_A}^{(A)} , \lambda_{l_B}^{(B)})     ,    
\end{split}  
\end{equation}
where 
\begin{equation} \label{F}
\begin{split}
&F(n, a, b) \equiv \sum_{k=0}^{n-1} \left(1- \cos \left( \frac{2\pi k}{n} \right) \right)^2 \frac{1}{1-a \cos \left( \frac{2\pi k}{n} \right) } \frac{1}{1-b \cos \left( \frac{2\pi k}{n} \right) }  .
\end{split}  
\end{equation}
We can perform explicitly the summation in (\ref{F}) and obtain (see Appendix \ref{F calculation}) 
\begin{equation} \label{F result}
\begin{split}
	&F(n, a, b) = n \frac{(1+p^2)(1+q^2)}{ 4pq  } 
	\Bigl[ 2- \frac{(1-p)(1+p^n)}{(1+p)(1-p^n)} - \frac{(1-q)(1+q^n)}{(1+q)(1-q^n)}  \\
	& +\frac{2}{(1+p)(1+q)(p-q)(1-pq)} \Bigl\{ \frac{1-p}{1-p^n} p (1-q)^2 (1+q) - \frac{1-q}{1-q^n} q (1-p)^2 (1+p) \\
	&-(p-q) (1+pq-(p+q)pq) \Bigr\} \Bigr]
\end{split}  
\end{equation}
where
\begin{equation}
\begin{split}
&p\equiv \rho(a)=\frac{1}{a} (1-\sqrt{1-a^2}), \, \, \, q\equiv \rho(b)=\frac{1}{b} (1-\sqrt{1-b^2}) .
\end{split}  
\end{equation}
From (\ref{F result}), for $n=1,2,3$ and $4$, we obtain  
\begin{equation} \label{F n=1}
\begin{split}
&F(n=1, a, b) =0
\end{split}  
\end{equation}
\begin{equation}
\begin{split}
&F(n=2, a, b) =4 \frac{(1+p^2)}{(1+p)^2 } \, \frac{(1+q^2)}{ (1+q)^2}
\end{split}  
\end{equation}
\begin{equation}
\begin{split}
&F(n=3, a, b) =\frac{9}{2} \frac{1+p^2}{1+p+p^2 } \,\frac{1+q^2}{1+q+q^2 }
\end{split}  
\end{equation}
\begin{equation}
\begin{split}
&F(n=4, a, b) =2 \left[ 1+ 2 \frac{(1+p^2)}{(1+p)^2 } \, \frac{(1+q^2)}{ (1+q)^2} \right] .
\end{split}  
\end{equation}
When $n=2,3$, $F(n, a, b)$ is a product of the function of $a$ and $b$ and $C_{AB}^{(n)} $ becomes,
\begin{equation}
\begin{split}
&C_{AB}^{(n)}=\tilde{C}^{(n)}_A \tilde{C}^{(n)}_B, ~~~(n=2,3)
\end{split}  
\end{equation}
where $\tilde{C}^{(n)}_{A(B)}$ is a function which is determined by the shape of A(B).
%In this case, $\tilde{C}^{(n)}_{A(B)}$ is a simple product of 
So, when $n=2,3$, $C^{(n)}_{AB}$ is not entangled, i.e. it is a simple product of functions each of which is determined by the shape of A(B). 
In general, $F(n, a, b)$ is not a product of the function of $a$ and $b$ and $C_{AB}^{(n)} $ is entangled.

Because $F(n, a, b)$ is an elementary function of $n$, its analytical continuation is trivial. 
So we can take $n \to 1$ limit in $C^{(n)}_{AB}$ in (\ref{C_{AB}}). 
From (\ref{F result}) and (\ref{F n=1}), we obtain
\begin{equation} \label{C_{AB}^{(n=1)}}
\begin{split}
&C^{(n=1)}_{AB}
=4  \sum_{i_A} \sum_{j_A} \sum_{l_A} \sum_{i_B} \sum_{j_B} \sum_{l_B} 
Z_{l_A i_A}^{(A)} Z_{l_A j_A}^{(A)} Z_{l_B i_B}^{(B)} Z_{l_B j_B}^{(B)}  \left.  \left( \frac{\partial }{\partial n } F(n, \lambda_{l_A}^{(A)}, \lambda_{l_B}^{(B)}) \right) \right|_{n=1}    ,    
\end{split}  
\end{equation}
where 
\begin{equation}
\begin{split}
&\left.  \left( \frac{\partial }{\partial n } F(n, a, b) \right) \right|_{n=1}
=\frac{1}{2} \frac{(1+p^2)(1+q^2)}{(1+p)(1+q)(p-q)(1-pq)} \left[ (1-p)(1+q)\ln p -  (1+p)(1-q)\ln q   \right] .
\end{split}  
\end{equation}
$C_{AB}^{(n=1)} $ is entangled.
The calculation of the matrix $Z$ and the eigenvalues $\lambda_i$ is simple matrix computation. 
So, we can compute $C^{(n=1)}_{AB}$ numerically. 
Note that (\ref{MI final}) is the mutual information of a free complex scalar field and the mutual information of a free real scalar field is a half of   (\ref{MI final}).

\section{Conclusion and discussions}

In this paper, we considered the (R\'{e}nyi) mutual information, $I^{(n)}(A,B)=S^{(n)}_A+S^{(n)}_{B}-S^{(n)}_{A \cup B}$, of distant compact spatial regions A and B in the vacuum state of a free scalar field. 
The distance r between A and B is much greater than their sizes $R_{A,B}$ and 	the  (R\'{e}nyi) mutual information behaves as $I^{(n)}(A,B) \sim C^{(n)}_{AB} \bra{0} \phi(r) \phi(0) \ket{0}^2$. 
%It is known that \cite{Ca1}, when $r \gg R_{A,B} $, the (R\'{e}nyi) mutual information behaves as 
%\begin{equation}
%I^{(n)}(A,B) \sim C^{(n)}_{AB} \bra{0} \phi(r) \phi(0) \ket{0}^2 ,
%\end{equation} 
%$I^{(n)}(A,B) \sim C^{(n)}_{AB} \left< \phi(r)\phi(0) \right>^2$ . 
%where $C^{(n)}_{AB}$ depends on the shapes of the regions A and B.  
We obtained the direct expression of $C^{(n)}_{AB}$ for arbitrary regions A and B. 
We performed the analytical continuation of $n$ and obtain the mutual information $I(A,B)=\lim_{n \to 1} I^{(n)}(A,B)$. 
When $n=2,3$, $C^{(n)}_{AB}$ is not entangled, i.e. it is a simple product of functions each of which is determined by the shape of A(B).
For general $n$, $C^{(n)}_{AB}$ is not a simple product of functions each of which is determined by the shape of A(B) and $C^{(n)}_{AB}$ is entangled.
For example, $C^{(n=1)}_{AB}$ is entangled when $n=1,4$.

The direct expression is useful for the numerical computation. 
By using the direct expression, we can compute directly $I(A,B)$ without computing $S_A, S_B$ and $S_{A \cup B}$ respectively, so it reduces significantly the amount of computation.

It is an interesting future problem to apply our direct expression to study the shape dependence of $C^{(n)}_{AB}$. 
For example, the corner contribution to mutual information in (2+1) dimension is an interesting problem. 
The corner contributions to entanglement entropy in (2+1) dimension are universal and have important information of the QFT \cite{CH 2007,CHL 2009,HT 2007,FM 2006}, however, the corner contribution to mutual information has not been studied well. 
Our method is useful for studying the corner contributions of mutual information.
It is also an interesting future problem to generalize our method to the entanglement negativity \cite{VW,CCT 2012}.

\acknowledgments
I would like to thank Tokiro Numasawa, Sotaro Sugishita, Tadashi Takayanagi, Kotaro Tamaoka, and Kento Watanabe for useful comments and discussions. 
I also thank the Yukawa Institute for Theoretical Physics at Kyoto University.
Discussions during the workshop YITP-T-19-03 "Quantum Information and String Theory 2019" were useful.
This work was supported by JSPS KAKENHI Grant Number JP19K14721.

\appendix
\section{Derivation of $0 \leq \lambda_i <1$} \label{lambda range proof}
We show that the range of the eigenvalues $\lambda_i$ of $Y$ in (\ref{definition lambda}) is $0 \leq \lambda_i <1$. 
%From (\ref{commutators}), $W$ and $W^{-1}$ are positive definite symmetric matrices. 
$A$ and $D$ in (\ref{definition A}) and (\ref{definition D}) are positive definite symmetric matrices because $W$ and $W^{-1}$ are positive definite symmetric matrices. 
So, $X=(A^{-1}+D)^{-1/2}$ in (\ref{definition X Y}) is a  positive definite symmetric matrix. 

In order to show that $Y$ is a positive semidefinite matrix, we use the following identity,
\begin{equation} \label{identity matirix}
\begin{split}
\begin{pmatrix}
1 & 0 \\ 
0 & 1
\end{pmatrix}
= 
\begin{pmatrix} 
A & B  \\
B^T & C 
\end{pmatrix} 
\begin{pmatrix} 
D & E  \\
E^T & F 
\end{pmatrix}
=
\begin{pmatrix} 
A D+BE^T & AE+BF  \\
B^T D +CE^T & B^{T}E+CF 
\end{pmatrix}.
\end{split}  
\end{equation}
From (\ref{identity matirix}), we obtain $DA-1=-EB^T$ and $B^T=-F^{-1}E^T A$.
Thus we rewrite $D-A^{-1}$ in $Y$ as
\begin{equation} \label{D-A^{-1}}
\begin{split}
D-A^{-1}=(DA-1)A^{-1} =-EB^TA^{-1}= E F^{-1} E^{T}.
\end{split}  
\end{equation}
Because $F^{-1}$ is a positive definite matrix and (\ref{D-A^{-1}}), 
$D-A^{-1}$ is a positive semidefinite matrix. 
Therefore, $Y=X (D-A^{-1}) X$ is a positive semidefinite matrix and we obtain $0\leq \lambda_i$.

Next we consider the upper bound of $\lambda_i$. 
We rewrite $1-Y$ as
\begin{equation} \label{1-Y}
\begin{split}
1-Y &= (A^{-1}+D)^{-1/2} \left[A^{-1}+D -(D-A^{-1})\right] (A^{-1}+D)^{-1/2} \\
    &= (A^{-1}+D)^{-1/2} 2A^{-1} (A^{-1}+D)^{-1/2}  .
\end{split}  
\end{equation}
Because $A^{-1}$ is a positive definite matrix and (\ref{1-Y}), 
$1-Y$ is a positive definite matrix and we obtain $ \lambda_i <1$. 
Therefore, we have shown $0 \leq \lambda_i <1$.

\section{The calculation of $F(n, a, b)$ in (\ref{F})} \label{F calculation}
We calculate the summation $F(n, a, b)$ in (\ref{F}) for $0 \leq a <1, ~ 0 \leq b <1  $. 
We expand $ \left(1- \cos \left( \frac{2\pi k}{n} \right) \right)^2$ in (\ref{F}) and rewrite $F(n, a, b)$ as
\begin{equation} \label{F sum 2}
\begin{split}
F(n, a, b) &= \frac{n}{ab} + \frac{1}{a} \left( 1-\frac{1}{b} \right) \sum_{k=0}^{n-1}  \frac{1}{1-b \cos \left( \frac{2\pi k}{n} \right) } +\frac{1}{b} \left( 1-\frac{1}{a} \right) \sum_{k=0}^{n-1}  \frac{1}{1-a \cos \left( \frac{2\pi k}{n} \right) } \\
&+  \left( 1-\frac{1}{a} \right)  \left( 1-\frac{1}{b} \right) \sum_{k=0}^{n-1}  \frac{1}{1-a \cos \left( \frac{2\pi k}{n} \right) }  \frac{1}{1-b \cos \left( \frac{2\pi k}{n} \right) } .
\end{split}  
\end{equation}
In order to calculate the summations in (\ref{F sum 2}), we use the following expansion,
\begin{equation} \label{expansion rho}
\begin{split}
\frac{1}{1-a \cos \theta} = (1+\rho (a)^2) \frac{1}{1-\rho (a) e^{i \theta}} \frac{1}{1-\rho (a) e^{- i \theta}} 
=  (1+\rho (a)^2) \sum_{p=0}^{\infty} \sum_{p'=0}^{\infty} \rho (a)^{p+p'}   e^{ i (p-p') \theta}
\end{split}  
\end{equation}
where
\begin{equation}
\begin{split}
&\rho (a) \equiv \frac{1}{a} (1-\sqrt{1-a^2}) ,
\end{split}  
\end{equation}
here $0 \leq \rho (a) <1 $ for $0\leq a <1$, and
\begin{equation}
\begin{split}
& a=\frac{2 \rho(a)}{1+\rho(a)^2} .
\end{split}  
\end{equation}
The expansion (\ref{expansion rho}) in the limit $a \to 1-$ was used in \cite{Ca1}.

\subsection{The calculation of $\sum_{k=0}^{n-1}  \frac{1}{1-a \cos \left( \frac{2\pi k}{n} \right) }$}
By using the expansion in (\ref{expansion rho}), we obtain
\begin{equation} \label{sum cos 1}
\begin{split}
& \sum_{k=0}^{n-1}  \frac{1}{1-a \cos \left( \frac{2\pi k}{n} \right) }=
(1+\rho (a)^2) \sum_{p=0}^{\infty} \sum_{p'=0}^{\infty} \rho (a)^{p+p'} \sum_{k=0}^{n-1}  e^{ i (p-p') \frac{2\pi k}{n}} .
\end{split}  
\end{equation}
We split the $p, p'$ summation into three parts,
\begin{equation} \label{sum p p'}
\begin{split}
& \sum_{p=0}^{\infty} \sum_{p'=0}^{\infty} 
=\left. \sum_{p'=0}^{\infty} \sum_{l=0}^{\infty} \right|_{p=p'+l} 
+ \left. \sum_{p=0}^{\infty} \sum_{l=0}^{\infty} \right|_{p'=p+l} 
-\left. \sum_{p=0}^{\infty}  \right|_{p'=p} ,
\end{split}  
\end{equation}
where we subtracted the $p=p'$ part to avoid double counting.   
%We substitute (\ref{sum p p'}) into (\ref{sum cos 1}) and obtain 
From (\ref{sum cos 1}) and (\ref{sum p p'}), we obtain
\begin{equation} \label{sum p p' k}
\begin{split}
&  \sum_{p=0}^{\infty} \sum_{p'=0}^{\infty} \rho (a)^{p+p'} \sum_{k=0}^{n-1}  e^{ i (p-p') \frac{2\pi k}{n}} \\
& = \sum_{p'=0}^{\infty} \sum_{l=0}^{\infty} \rho^{2p'+l} \sum_{k=0}^{n-1} e^{ i l \frac{2\pi k}{n}} 
+  \sum_{p=0}^{\infty} \sum_{l=0}^{\infty} \rho^{2p+l} \sum_{k=0}^{n-1} e^{- i l \frac{2\pi k}{n}} 
- \sum_{p=0}^{\infty} \rho^{2p} \sum_{k=0}^{n-1} 1 \\
& = n \sum_{p'=0}^{\infty} \sum_{j=0}^{\infty} \rho^{2p'+nj} 
+ n \sum_{p=0}^{\infty} \sum_{j=0}^{\infty} \rho^{2p+nj} 
-n \sum_{p=0}^{\infty} \rho^{2p} \\
& = \frac{2n}{(1-\rho^2)(1-\rho^n)} - \dfrac{n}{1-\rho^2} = \frac{n}{1-\rho^2} \cdot \frac{1+\rho^n}{1-\rho^n}  
\end{split}  
\end{equation}
where we have used
\begin{equation} \label{delta function}
\begin{split}
\sum_{k=0}^{n-1} e^{ i l \frac{2\pi k}{n}} =
n \delta_{l,nj} ~~  (j=0,1,2,\cdots)   .
\end{split}  
\end{equation}
We substitute (\ref{sum p p' k}) into (\ref{sum cos 1}) and obtain 
\begin{equation} \label{sum cos 1 result}
\begin{split}
& \sum_{k=0}^{n-1}  \frac{1}{1-a \cos \left( \frac{2\pi k}{n} \right) }
=(1+\rho (a)^2) \frac{n}{1-\rho(a)^2} \cdot \frac{1+\rho(a)^n}{1-\rho(a)^n}  .
\end{split}  
\end{equation}

\subsection{The calculation of $\sum_{k=0}^{n-1}  \frac{1}{1-a \cos \left( \frac{2\pi k}{n} \right) } 	\frac{1}{1-b \cos \left( \frac{2\pi k}{n} \right) }$}
In the same way as above, by using the expansion in (\ref{expansion rho}), we obtain 
\begin{equation} \label{sum cos 2}
\begin{split}
& \sum_{k=0}^{n-1}  \frac{1}{1-a \cos \left( \frac{2\pi k}{n} \right) } 
\frac{1}{1-b \cos \left( \frac{2\pi k}{n} \right) } \\
&=(1+\rho (a)^2) (1+\rho (b)^2) 
\sum_{k=0}^{n-1}
\sum_{p=0}^{\infty} \sum_{p'=0}^{\infty} \rho (a)^{p+p'}   e^{ i (p-p') \frac{2\pi k}{n}}  
\sum_{q=0}^{\infty} \sum_{q'=0}^{\infty} \rho (b)^{q+q'}   e^{ i (q-q') \frac{2\pi k}{n}}  .
\end{split}  
\end{equation}
From (\ref{sum p p'}), we can rewrite the summations in (\ref{sum cos 2}) as 
\begin{equation} \label{sum p p' q q'}
\begin{split}
& \sum_{p=0}^{\infty} \sum_{p'=0}^{\infty}  \sum_{q=0}^{\infty} \sum_{q'=0}^{\infty} \\
& =\left( \left. \sum_{p'=0}^{\infty} \sum_{l=0}^{\infty} \right|_{p=p'+l} 
+ \left. \sum_{p=0}^{\infty} \sum_{l=0}^{\infty} \right|_{p'=p+l} 
-\left. \sum_{p=0}^{\infty}  \right|_{p'=p} \right)
\left( \left. \sum_{q'=0}^{\infty} \sum_{m=0}^{\infty} \right|_{q=q'+m} 
+ \left. \sum_{q=0}^{\infty} \sum_{m=0}^{\infty} \right|_{q'=q+m} 
-\left. \sum_{q=0}^{\infty}  \right|_{q'=q} \right) .
\end{split}  
\end{equation}
We substitute (\ref{sum p p' q q'}) into (\ref{sum cos 2}) and obtain
\begin{equation} \label{sum cos 2-2}
\begin{split}
& \frac{1}{(1+\rho (a)^2)(1+\rho (b)^2)} \sum_{k=0}^{n-1}  \frac{1}{1-a \cos \left( \frac{2\pi k}{n} \right) } 
\frac{1}{1-b \cos \left( \frac{2\pi k}{n} \right) } \\
&=\sum_{k=0}^{n-1} 
\left[2\left( \sum_{p'=0}^{\infty} \sum_{l=0}^{\infty}  \sum_{q'=0}^{\infty} \sum_{m=0}^{\infty} \rho(a)^{2p'+l} \rho(b)^{2q'+m} e^{i(l+m)2\pi k/n} \right. \right. \\
& +\sum_{p'=0}^{\infty} \sum_{l=0}^{\infty}  \sum_{q=0}^{\infty} \sum_{m=0}^{\infty} \rho(a)^{2p'+l} \rho(b)^{2q+m} e^{i(l-m)2\pi k/n} \\
& -\sum_{p'=0}^{\infty} \sum_{l=0}^{\infty}  \sum_{q=0}^{\infty} 
\rho(a)^{2p'+l} \rho(b)^{2q} e^{il 2\pi k/n} \\
&  \left. -\sum_{p=0}^{\infty}    \sum_{q'=0}^{\infty}  \sum_{m=0}^{\infty}
\rho(a)^{2p} \rho(b)^{2q'+m} e^{im 2\pi k/n} \right)  \\
& \left. +\sum_{p=0}^{\infty} \sum_{q=0}^{\infty} \rho(a)^{2p} \rho(b)^{2q} \right] .
\end{split}  
\end{equation}
The last term in (\ref{sum cos 2-2}) can be evaluated as
\begin{equation} \label{last term}
\begin{split}
\sum_{k=0}^{n-1} \sum_{p=0}^{\infty} \sum_{q=0}^{\infty} \rho(a)^{2p} \rho(b)^{2q} 
= n \frac{1}{1-\rho(a)^2}  \frac{1}{1-\rho(b)^2} .
\end{split}  
\end{equation}
The third term in (\ref{sum cos 2-2}) can be evaluated as
\begin{equation} \label{third term}
\begin{split}
\sum_{k=0}^{n-1} \sum_{p'=0}^{\infty} \sum_{l=0}^{\infty}  \sum_{q=0}^{\infty} 
\rho(a)^{2p'+l} \rho(b)^{2q} e^{il 2\pi k/n} 
&= n \sum_{p'=0}^{\infty} \sum_{j=0}^{\infty}  \sum_{q=0}^{\infty} 
\rho(a)^{2p'+n j} \rho(b)^{2q}  \\
&= n \frac{1}{1-\rho(a)^2} \frac{1}{1-\rho(a)^n} \frac{1}{1-\rho(b)^2} .
\end{split}  
\end{equation}
The fourth term in (\ref{sum cos 2-2}) is obtained by interchanging $a$ and $b$ in the third term in (\ref{sum cos 2-2}). 

We perform the $p'$ and $q$ summations in the second term in (\ref{sum cos 2-2}) and obtain
\begin{equation} \label{sum k p' l q m}
\begin{split}
&\sum_{k=0}^{n-1} \sum_{p'=0}^{\infty} \sum_{l=0}^{\infty}  \sum_{q=0}^{\infty} \sum_{m=0}^{\infty} \rho(a)^{2p'+l} \rho(b)^{2q+m} e^{i(l-m)2\pi k/n} \\
&= \frac{1}{1-\rho(a)^2} \frac{1}{1-\rho(b)^2} \sum_{k=0}^{n-1} \sum_{l=0}^{\infty} \sum_{m=0}^{\infty}  \rho(a)^{l} \rho(b)^{m} e^{i(l-m)2\pi k/n} .
\end{split}  
\end{equation}
By using (\ref{sum p p'}), we obtain
\begin{equation} \label{sum k l m result}
\begin{split}
& \sum_{k=0}^{n-1} \sum_{l=0}^{\infty} \sum_{m=0}^{\infty}  \rho(a)^{l} \rho(b)^{m} e^{i(l-m)2\pi k/n} \\
&= \sum_{k=0}^{n-1} \left[ \sum_{l=0}^{\infty} \sum_{\alpha=0}^{\infty} \rho(a)^{l} \rho(b)^{l+\alpha} e^{-i\alpha 2\pi k/n} 
+ \sum_{m=0}^{\infty} \sum_{\alpha=0}^{\infty} \rho(a)^{m+\alpha} \rho(b)^{m} e^{i\alpha 2\pi k/n} 
- \sum_{l=0}^{\infty}  \rho(a)^{l} \rho(b)^{l}  \right] \\
&=n \sum_{l=0}^{\infty} \sum_{j=0}^{\infty}  \rho(a)^{l} \rho(b)^{l+nj} 
+n \sum_{m=0}^{\infty} \sum_{j=0}^{\infty}  \rho(a)^{m+nj} \rho(b)^{m} 
-n \frac{1}{1-\rho(a)\rho(b)}  \\
&=\frac{n}{1-\rho(a)\rho(b)}\left[\frac{1}{1-\rho(b)^n} + \frac{1}{1-\rho(a)^n} - 1  \right].
\end{split}  
\end{equation}
Thus, we substitute (\ref{sum k l m result}) into (\ref{sum k p' l q m}) and obtain the second term in (\ref{sum cos 2-2})
\begin{equation} \label{second term}
\begin{split}
&\sum_{k=0}^{n-1} \sum_{p'=0}^{\infty} \sum_{l=0}^{\infty}  \sum_{q=0}^{\infty} \sum_{m=0}^{\infty} \rho(a)^{2p'+l} \rho(b)^{2q+m} e^{i(l-m)2\pi k/n} \\
&= \frac{1}{1-\rho(a)^2} \frac{1}{1-\rho(b)^2} 
\frac{n}{1-\rho(a)\rho(b)}\left[\frac{1}{1-\rho(a)^n} + \frac{1}{1-\rho(b)^n} - 1  \right] .
\end{split}  
\end{equation}

We perform the $p'$ and $q'$ summations in the first term in (\ref{sum cos 2-2}) and obtain
\begin{equation} \label{sum k p' l q' m}
\begin{split}
&\sum_{k=0}^{n-1} \sum_{p'=0}^{\infty} \sum_{l=0}^{\infty}  \sum_{q'=0}^{\infty} \sum_{m=0}^{\infty} \rho(a)^{2p'+l} \rho(b)^{2q'+m} e^{i(l+m)2\pi k/n}  \\
&= \frac{1}{1-\rho(a)^2} \frac{1}{1-\rho(b)^2} \sum_{k=0}^{n-1} \sum_{l=0}^{\infty} \sum_{m=0}^{\infty}  \rho(a)^{l} \rho(b)^{m} e^{i(l+m)2\pi k/n}
\end{split}  
\end{equation}
By using (\ref{delta function}), we obtain
\begin{equation} \label{sum k l m result 2}
\begin{split}
& \sum_{k=0}^{n-1} \sum_{l=0}^{\infty} \sum_{m=0}^{\infty}  \rho(a)^{l} \rho(b)^{m} e^{i(l+m)2\pi k/n} 
= n \sum_{j=0}^{\infty} \sum_{l=0}^{nj} \rho(a)^l  \rho(b)^{nj-l} \\
&=n \sum_{j=0}^{\infty} \rho(b)^{nj} \left(\frac{1-(\rho(a)/\rho(b))^{nj+1}}{1-(\rho(a)/\rho(b))} \right) =\frac{n}{\rho(b)-\rho(a)} \left[\frac{\rho(b)}{1-\rho(b)^n} -\frac{\rho(a)}{1-\rho(a)^n} \right] .
\end{split}  
\end{equation}
We substitute (\ref{sum k l m result 2}) into (\ref{sum k p' l q' m}) and obtain the first term in (\ref{sum cos 2-2})
\begin{equation} \label{first term}
\begin{split}
&\sum_{k=0}^{n-1} \sum_{p'=0}^{\infty} \sum_{l=0}^{\infty}  \sum_{q'=0}^{\infty} \sum_{m=0}^{\infty} \rho(a)^{2p'+l} \rho(b)^{2q'+m} e^{i(l+m)2\pi k/n}  \\
&= \frac{1}{1-\rho(a)^2} \frac{1}{1-\rho(b)^2} 
\frac{n}{\rho(b)-\rho(a)} \left[\frac{\rho(b)}{1-\rho(b)^n} -\frac{\rho(a)}{1-\rho(a)^n} \right] .
\end{split}  
\end{equation}

Finally, we substitute (\ref{last term}), (\ref{third term}), (\ref{second term}) and (\ref{first term}) into (\ref{sum cos 2-2}) and obtain
\begin{equation} \label{sum cos 2 result}
\begin{split}
& \frac{1}{(1+\rho (a)^2)(1+\rho (b)^2)} \sum_{k=0}^{n-1}  \frac{1}{1-a \cos \left( \frac{2\pi k}{n} \right) } 
\frac{1}{1-b \cos \left( \frac{2\pi k}{n} \right) } \\
&= 2\left[ \frac{1}{1-\rho(a)^2} \frac{1}{1-\rho(b)^2} 
\frac{n}{\rho(b)-\rho(a)} \left( \frac{\rho(b)}{1-\rho(b)^n} -\frac{\rho(a)}{1-\rho(a)^n} \right) \right. \\ 
&+ \frac{1}{1-\rho(a)^2} \frac{1}{1-\rho(b)^2} 
\frac{n}{1-\rho(a)\rho(b)}\left(\frac{1}{1-\rho(a)^n} + \frac{1}{1-\rho(b)^n} - 1  \right) \\
&\left. -  n \frac{1}{1-\rho(a)^2}  \frac{1}{1-\rho(b)^2} \left(\frac{1}{1-\rho(a)^n}+\frac{1}{1-\rho(b)^n} \right)  \right] \\
&+n \frac{1}{1-\rho(a)^2}  \frac{1}{1-\rho(b)^2}  \\
&= 2n \frac{1}{1-\rho(a)^2}  \frac{1}{1-\rho(b)^2} \frac{1}{\rho(a)-\rho(b)} \frac{1}{1-\rho(a)\rho(b)} \\
&\times \left[\frac{\rho(a)(1-\rho(b)^2)}{1-\rho(a)^n} -\frac{\rho(b)(1-\rho(a)^2)}{1-\rho(b)^n} 
-\frac{1}{2} (\rho(a)-\rho(b))(1+\rho(a)\rho(b))  \right] .
\end{split}  
\end{equation}

We substitute (\ref{sum cos 1 result}) and (\ref{sum cos 2 result}) into (\ref{F sum 2}) and obtain %$F(n, a, b)$ for $0 \leq a <1, ~ 0 \leq b <1  $
\begin{equation} \label{F in appendix B}
\begin{split}
&F(n, a, b) = n \frac{(1+p^2)(1+q^2)}{ 4pq  } 
 \Bigl[ 2- \frac{(1-p)(1+p^n)}{(1+p)(1-p^n)} - \frac{(1-q)(1+q^n)}{(1+q)(1-q^n)}  \\
& +\frac{2}{(1+p)(1+q)(p-q)(1-pq)} \Bigl\{ \frac{1-p}{1-p^n} p (1-q)^2 (1+q) - \frac{1-q}{1-q^n} q (1-p)^2 (1+p) \\
&-(p-q) (1+pq-(p+q)pq) \Bigr\} \Bigr]
\end{split}  
\end{equation}
where $0 \leq a <1, ~ 0 \leq b <1  $ and
\begin{equation}
\begin{split}
&p\equiv \rho(a)=\frac{1}{a} (1-\sqrt{1-a^2}), \, \, \, q\equiv \rho(b)=\frac{1}{b} (1-\sqrt{1-b^2}).
\end{split}  
\end{equation}
Thus (\ref{F in appendix B}) is equal to (\ref{F}).

%the end of appendix B

%We would like to thank Masahiro Nozaki, Shinsei Ryu and Erik Tonni for useful discussions.
%NS and TT are supported by JSPS Grant-in-Aid for Scientific
%Research (B) No.25287058 and JSPS Grant-in-Aid for Challenging
%Exploratory Research No.24654057. TT is also
%supported by World Premier International
%Research Center Initiative (WPI Initiative) from the Japan Ministry
%of Education, Culture, Sports, Science and Technology (MEXT).

%This is the most common positions for acknowledgments. A macro is
%available to maintain the same layout and spelling of the heading.

%\paragraph{Note added.} This is also a good position for notes added
%after the paper has been written.


\end{document}